# Machine learning assisted screening of metal binary alloys for anode materials

Xingyue Shi, Linming Zhou, Yuhui Huang, Yongjun Wu, Zijian Hong

School of Materials Science and Engineering, Zhejiang University, Hangzhou, Zhejiang 310027, China

## Abstract

In the dynamic and rapidly advancing battery field, alloy anode materials are a focal point due to their superior electrochemical performance. Traditional screening methods are inefficient and time-consuming. Our research introduces a machine learning-assisted strategy to expedite the discovery and optimization of these materials. We compiled a vast dataset from the MP and AFLOW databases, encompassing tens of thousands of alloy compositions and properties. Utilizing a CGCNN, we accurately predicted the potential and specific capacity of alloy anodes, validated against experimental data. This approach identified approximately 120 low potential and high specific capacity alloy anodes suitable for various battery systems including Li, Na, K, Zn, Mg, Ca, and Al-based. Our method not only streamlines the screening of battery anode materials but also propels the advancement of battery material research and innovation in energy storage technology.

## Introduction

Battery is one of the most important energy storage technologies and plays a key role in new energy vehicles and electronic devices[1]. Anode materials are a critical component of batteries, directly impacting their energy density, cycle life, and safety [2]. Traditional anode materials like graphite have a low theoretical capacity (372 mAh $g^{-1}$), which doesn't meet the demand for high energy density batteries [3]. Silicon and germanium anode materials have high theoretical specific capacity (Si: 4200 mAh $g^{-1}$, Ge: 1600 mAh $g^{-1}$) [4], but they suffer from serious capacity degradation and undergo large volume expansions during charging and discharging, leading to structural damage and electrode failure. Silicon-based anode materials can form an unstable solid electrolyte interphase (SEI) layer, reducing Coulombic Efficiency [5]. Metal anodes face challenges such as safety concerns due to dendrite growth and low Coulombic Efficiency from volume expansion. Lead anodes, due to their high density, are not suitable for high-energy-density batteries.

In contrast, alloy anode materials have attracted much attention due to their higher energy density, specific capacity, and good rate capability [6]. Alloy materials can also effectively inhibit the growth of dendrites by lowering the embedding potential and optimizing the electrode structure, thus improving the safety and cycling stability of batteries. Therefore, the development of new high-capacity and high-stability anode materials is crucial for improving battery performance. Alloy-type anode materials, such as Sb, Bi, and Sn, show high theoretical capacities, but still face many challenges in practical applications, and further in-depth research is needed to overcome these challenges.

In recent years, data-driven machine learning techniques have emerged as one of the hotspots of research in the field of materials science and engineering. It can quickly identify material performance, predict material behavior and optimize material design by analyzing and processing large amounts of data, thus significantly shortening the research and development cycle of new materials and improving research and development efficiency. In the battery field, the application of machine learning is particularly important [7]. The development of battery materials is a complex and time-consuming process, involving a combination of multiple chemical compositions and physical structures. The traditional trial-and-error method is not only time-consuming and labor-intensive, but also costly. Machine learning techniques can learn from massive amounts of experimental data, discover the correlation between material performance and structure, and predict the performance of novel materials in batteries, thus accelerating the discovery and optimization of new materials. Current applications of machine learning in batteries include predicting battery life [8], electrode material design [9], and electrolyte optimization [9a, 10].

Our work utilizes a Crystal Graph Convolutional Neural Network (CGCNN)[11] to screen candidate anode materials for seven common types of batteries (Li, Na, K, Zn, Mg, Ca, and Al batteries) from tens of thousands of binary alloy compounds by examining the formation energy, potential, and specific capacity, providing new ideas for the design of battery electrode materials.

We seek superior anode materials by evaluating potential and specific capacity. The ideal anodes should have low formation energy, low potential, and high specific capacity. A low formation energy in anode materials means that ions within the material can quickly and effectively take part in electrochemical reactions during battery charging and discharging, improving the battery's rate performance. The ability to readily accept and release ions during these processes further increases efficiency. A lower potential increases the potential difference between the cathode and anode electrodes, boosting the battery's discharge voltage and, as a result, its energy density [9]. A high specific capacity allows for the storage of more charge and the release of more energy, prolonging working time and enhancing energy density. Additionally, it reduces the weight and volume of the battery, improving its portability and applicability.

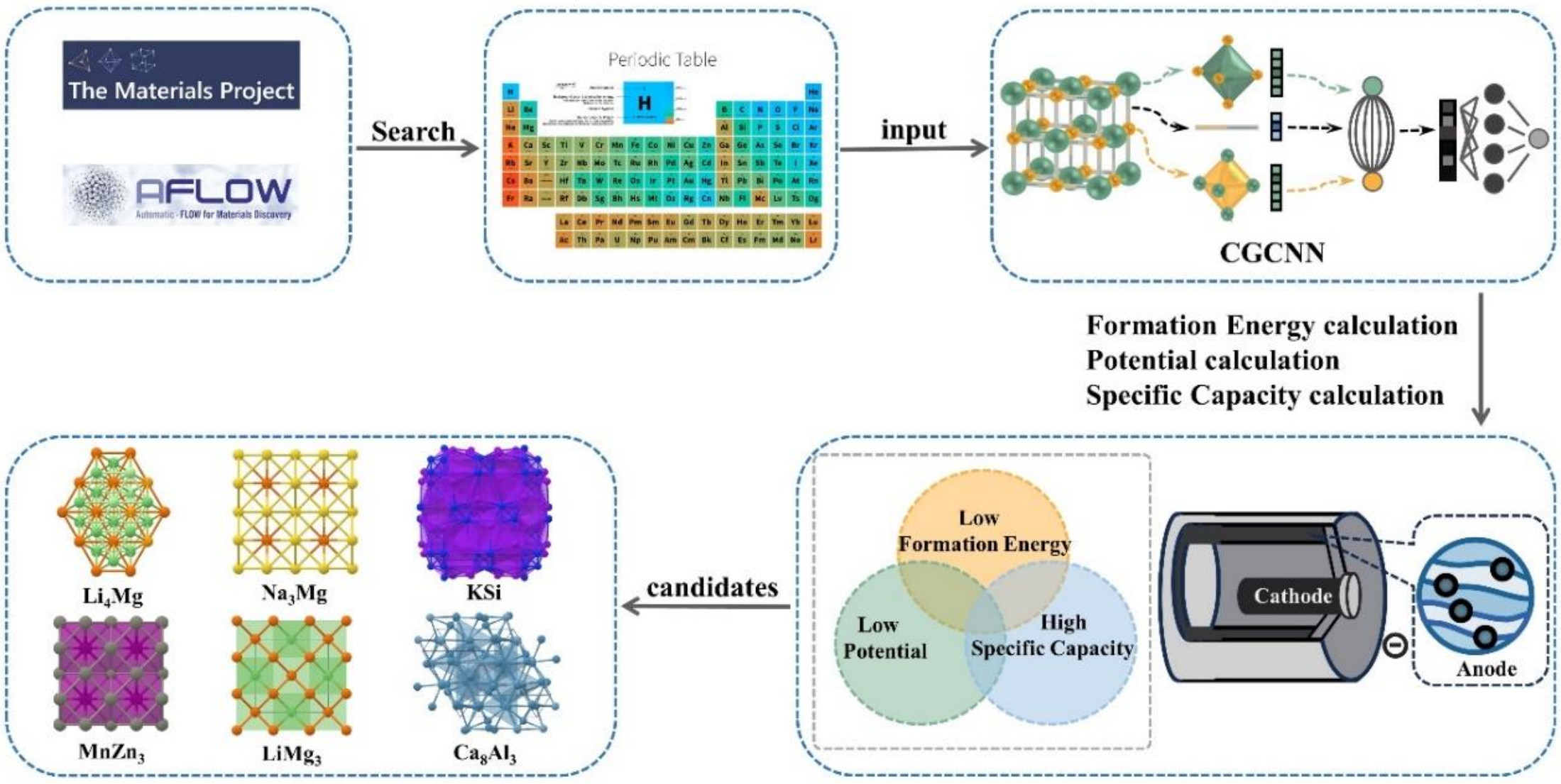

**Figure 1.** Flowchart of the machine learning-assisted screening of alloy anode materials. The dataset is obtained by combining the periodic table of elements using two databases, MP (Materials Project)[12] and AFLOW (Automatic FLOW for Materials Discovery [13]),

and CGCNN is used to obtain the formation energy, potential and specific capacity, and finally the candidate electrodes are screened.

The framework diagram of our work is shown in **Figure 1**. We obtained the energy as well as structural information of 12,405 binary alloys containing all metallic elements as well as common non-metallic elements (Si, P, C) forming the electrodes from MP using API (Pymatgen application programming interface). In addition, energy data and structural information of about 60,000 compounds were obtained from the AFLOW materials database. The energy data obtained from the two databases were used as the datasets for training the CGCNN models, where the ratio of training, validation, and test sets were 60%, 20%, and 20%, respectively, to obtain the two best models, MP-CGCNN and AFLOW-CGCNN [14] . Then we used AFLOW-CGCNN to predict the energy of the binary alloys obtained by MP, and obtained the Energy-ML.

We use the Gibbs free energy to calculate the potential. From the classical electrochemical theory, the reaction equation can be expressed as:

$$aM + bX^{n+} + nbe^{-} = M_aX_b \tag{1}$$

Here $M$ denotes the binary alloy and $X^{n+}$ denotes the ion that provides the charge in a common battery.

From the Gibbs free energy change $\Delta G$, we can calculate the electrochemical potential $\varphi$ of the cell:

$$\Delta G = G_{M_aX_b} - (aG_M + bG_X) \tag{2}$$

$$\varphi = -\Delta G/(2naF) \tag{3}$$

where $G_{M_aX_b}$ is the Gibbs free energy of the product $M_aX_b$, $G_M$ and $G_X$ are the Gibbs free energies of $M$ and $X$, respectively, $F$ is the Faraday's constant, and $2na$ is the number of electrons transferred in the reaction.

Based on Eqs. (2)(3), we calculated the potentials $\varphi_{MP}$ and $\varphi_{ML}$, respectively, and calculated $\varphi_{mix}$ [14]:

$$\varphi_{mix} = \sqrt{\varphi_{MP}\ \varphi_{ML}} \tag{4}$$

Theoretical Specific Capacity (TSC) is the amount of power that a battery material can provide per gram of material under ideal conditions, usually measured in milliampere hours per gram (mAh $g^{-1}$). This value is calculated based on the chemistry of the material and Faraday's Law. We take the following formula to calculate the specific capacity:

$$C = \frac{nF}{m_M} \tag{5}$$

$n$ is the number of electrons transferred by the ions involved in the reaction, $F$ is Faraday's constant, and $m_M$ is the molar mass of the binary alloy compound. Finally, we identify seven candidate electrode materials for common batteries based on low potential and high specific capacity.

## Results and Discussion

Since the potentials used in this study are derived by calculating the Gibbs free energy, we first need to explore the effect of energy representation on the CGCNN model. **Figure 2** shows how well the energy data obtained from the two material databases, MP and AFLOW, behave on the CGCNN model, where (a)(c) shows the scatter plots of the distribution of the energy values obtained from the databases versus the predicted values from the CGCNN model, and (b)(d) shows the trend of the Mean Absolute Error (MAE) as Epoch increases. These four plots allow a comprehensive evaluation of the performance of the energies obtained from the two databases on the CGCNN.

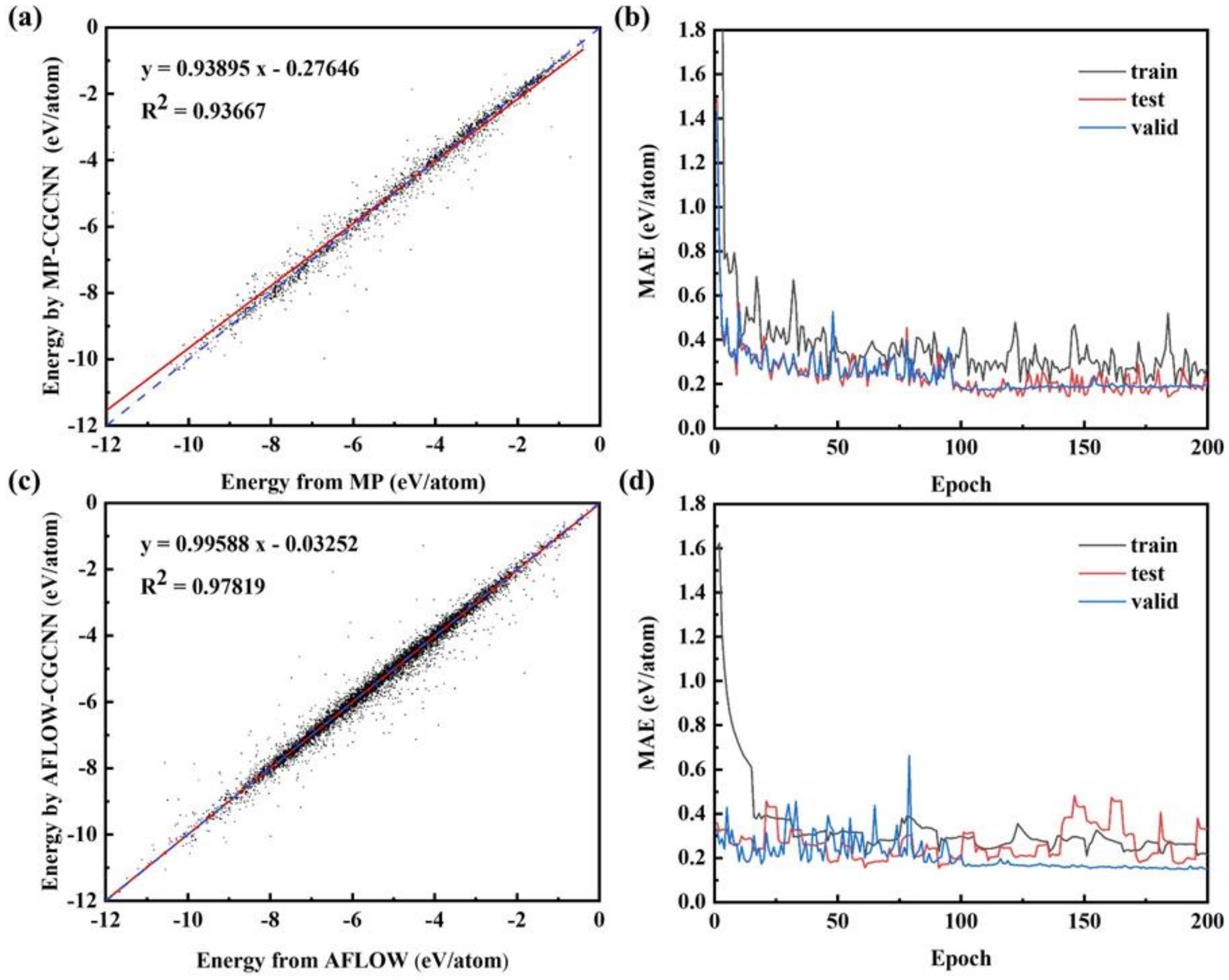

**Figure 2.** Performance of Gibbs free energy on CGCNN model from two databases. (a) Scatter plot of Gibbs free energy performance on CGCNN from MP database. (b) Plot of MAE with trained Epoch for training, test and validation sets. (c) Scatterplot of Gibbs free energy performance on CGCNN from AFLOW database. (d) Plot of MAE with trained Epoch for training, test and validation sets.

Analyzing **Figure 2**, it can be found that there is a strong linear correlation between the energy data extracted from the two material databases and the energy data trained by the CGCNN model. In the first 25 Epochs, the MAE decreases sharply and the $R^2$ (coefficients of determination) for both cases are close to 1. We set 200 Epochs, and we can see that after the 100th Epoch, the MAE stabilizes and fluctuates up and down around a value of 0.15 eV/atom, indicating that the model has converged. This result confirms that the energy not only performs well on the CGCNN model, but also has high accuracy.

The computational results of the potentials, calculated using the Gibbs free energy, are depicted in **Figure 3 and Figure S1**. It is evident that the majority of the calculated potentials exhibit positive values, with only approximately one-fifth yielding negative results. A deeper analysis indicates that, despite the variations in the specific values of $\varphi_{MP}$ and $\varphi_{ML}$ and their quantitative distributions, both datasets demonstrate a similar trend in their overall distribution. There are more types of lithium alloy anodes than sodium anode alloys, and the potential is relatively lower. The lithium alloy anode is mainly concentrated in about 0.2 V, while the potential of the sodium alloy anode is evenly distributed between 0 - 0.8 V. In contrast, there are fewer types of potassium and magnesium alloy anodes. Magnesium alloy anodes are primarily distributed between 0 V - 0.4 V, and potassium alloy potentials range from -1.0 V - 2.0 V.

To harmonize the potential data from the two datasets and enhance the precision of our calculations, we meticulously cleaned the data, retaining only those materials with negative potentials where both $\varphi_{MP}$ and $\varphi_{ML}$ were positive. Upon this refined dataset, we computed $\varphi_{mix}$ and visualized its distribution. Examination of the graphs reveals that the majority of $\varphi_{mix}$ values cluster within the 0 - 0.5 V. This distribution closely aligns with both our experimental findings and prior expectations, thereby furnishing robust data support for our investigation.

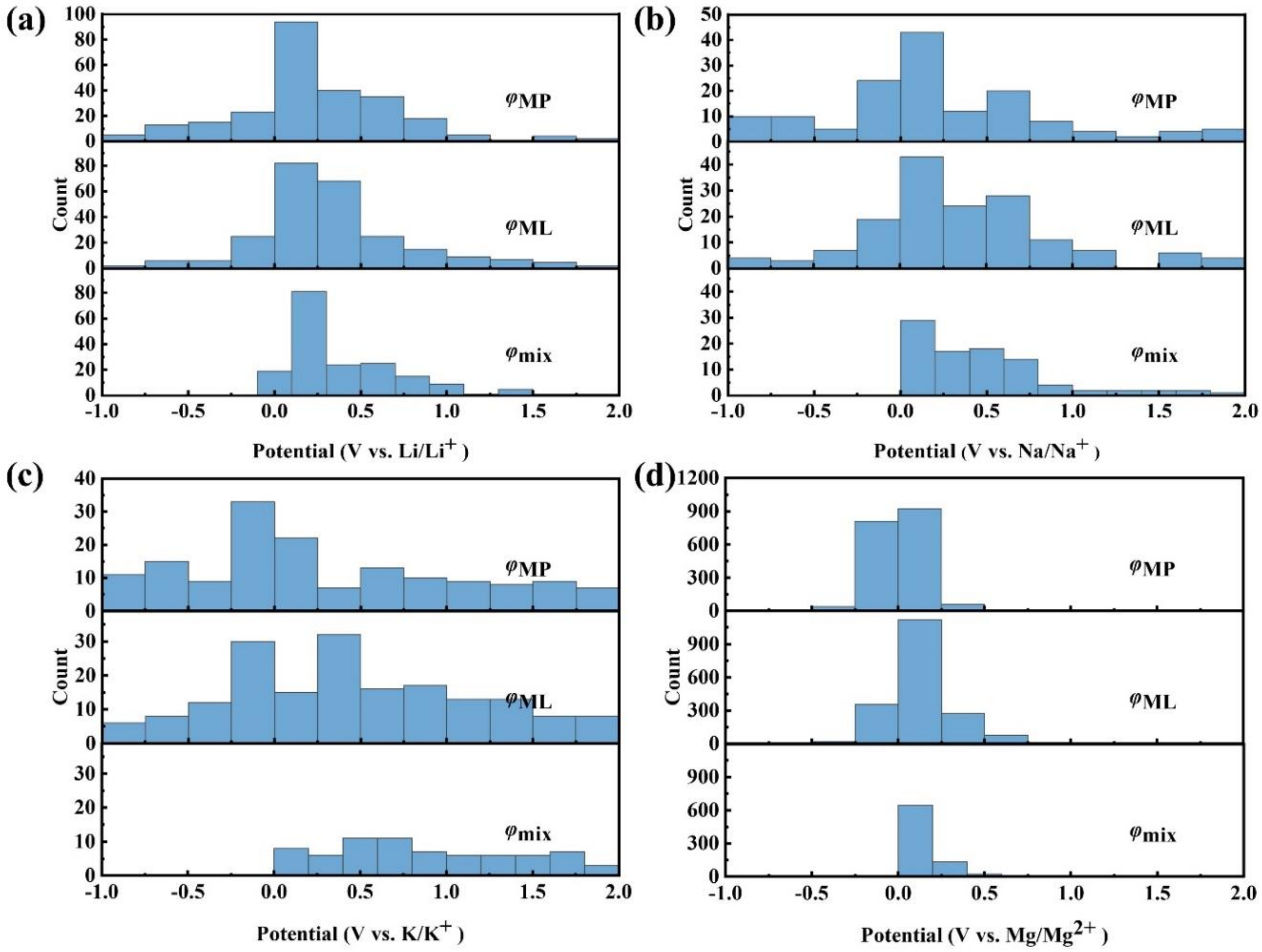

**Figure 3.** Calculated potentials and the quantitative distribution of the potentials. (a)(b)(c)(d) are all made up of three small figures, from top to bottom: Statistical distributions of potentials calculated from energies from the MP database ($\varphi_{MP}$); Potentials calculated from energies predicted by CGCNN-ML ($\varphi_{ML}$); Statistical distribution of $\varphi_{mix}$ ($\varphi_{mix} = \sqrt{\varphi_{MP}\,\varphi_{ML}}$).

This finding not only validates the effectiveness of our research methodology but also lays a solid foundation for subsequent research work. By comparing theoretical calculations with experimental data, we can more accurately grasp the electrochemical behavior of anode materials, thereby providing more precise guidance for battery design and optimization.

To further bolster the reliability of the candidate anode materials we have presented, we have meticulously compared the calculated potential values with experimental data from a variety of alloy anodes documented in published research. These alloy anodes encompass $Li_{22}Si_5$ [15], $Li_3Mg$ [6a], $Li_{15}Ge_4$ [16], $Li_{22}Sn_5$ [17], LiAl [18], $Li_2Ga$ [19], $Li_3Sb$ [20], LiZn [21], $Li_3Bi$ [20], $Li_5In_4$ [22], $Na_{15}Sn_4$ [23], $Na_2Se$ [24], $Na_3Sb$ [25], $Na_9Sn_4$ [23], $Na_2In$ [26], $Na_3Bi$ [27], NaGe [28], NaBi [27], $K_4P_3$ [29], $K_3Sb$ [30], $K_3Bi$ [31], KSn [32], KSb [33], $KSb_2$ [33], $Mg_2Sn$ [34], $Mg_3Sb_2$ [34], MgIn [35], $Mg_3Bi_2$ [34, 36], $Mg_2Ga_5$ [37], among others. As illustrated in **Figure 4**, we added error bars to the calculated potential values to better show the uncertainties in the potential calculations. Our rigorous comparison revealed a strong agreement between the experimental and calculated potential values, providing a solid empirical foundation for our research results.

Despite our extensive research, we have observed a notable paucity, or even a void, in studies concerning the anodes of alloys formed from other active metals such as Na, K, and Ca. This phenomenon is primarily due to the excessively strong chemical reactivity of these active metals, which poses significant challenges in experimental research. Consequently, there is an urgent need for further exploration and innovation in the study of these active metal alloy anodes, with the aim of bridging the existing research gaps and fostering the profound advancement of related scientific domains.

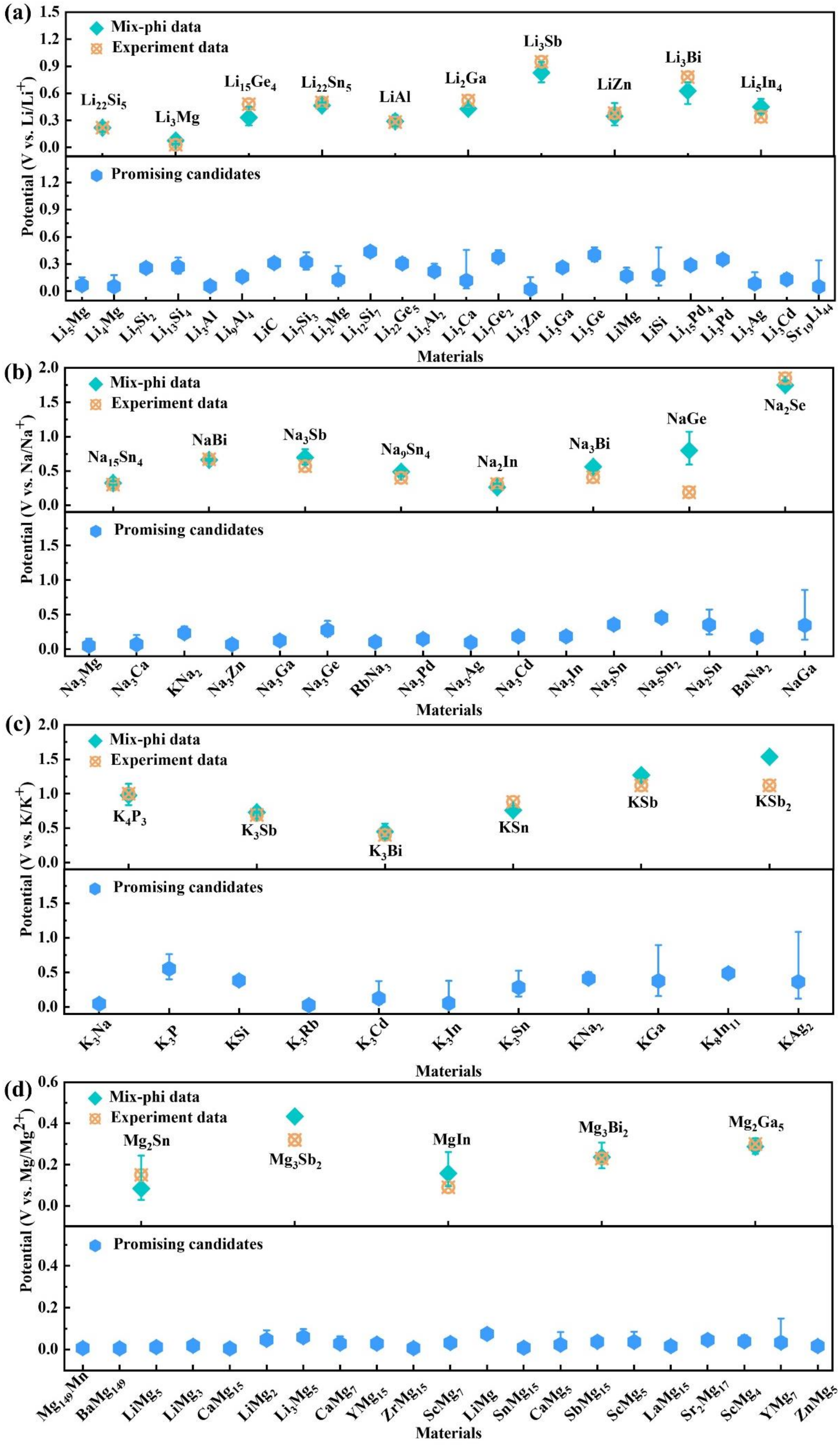

**Figure 4.** Comparison of calculated and experimental potential values for alloy anodes in Li, Na, K, and Mg batteries, corresponding candidate alloy anodes and their calculated potentials, with error bars added to show uncertainties. The candidate anodes, from left to right, have progressively decreasing theoretical specific capacities.

Presently, some studies have conducted initial explorations on alloy anodes of Li and Na. Through screening and prediction, we have identified dozens of potential alloy anode materials, such as $Li_4Mg$, $Li_7Si_2$, $Li_3Al$, $Na_3Mg$, $Na_3Ca$, etc. The materials mentioned exhibit excellent specific capacities and low electrode potentials. It's important to note that these alloy anode materials follow a certain pattern in terms of the added elements, with a majority of them being focused on elements like Si, Sn, Sb, In, and Bi. This suggests that these elements have the potential to enhance the performance of alloy anodes of active metals and warrant further in-depth research. However, their actual performance still needs to be verified through experiments to address challenges such as poor cyclic stability and significant volume expansion.

In comparison, research on potassium (K) and magnesium (Mg) alloy anodes is relatively lacking, with few relevant literature reports, mainly focusing on a few elements like Stannum (Sn) and Antimony (Sb). Notably, indium (In) has attracted attention as a potential potassium alloy anode material due to its low potential and good stability. For Mg-based alloy anodes, although numerous candidate materials have been predicted, further screening and optimization are required to achieve ideal materials with both high capacity and stability.

Furthermore, research on alloy anodes of other metals such as zinc (Zn), calcium (Ca), and aluminum (Al) is even more limited, with some areas remaining unexplored.

This is mainly due to various challenges in their usage environments or applications, including fewer advancements in alloying, or the high reactivity of the metals, making it difficult to study and prepare alloy anode materials. Therefore, theoretical calculations are currently used to predict their electrochemical performance.

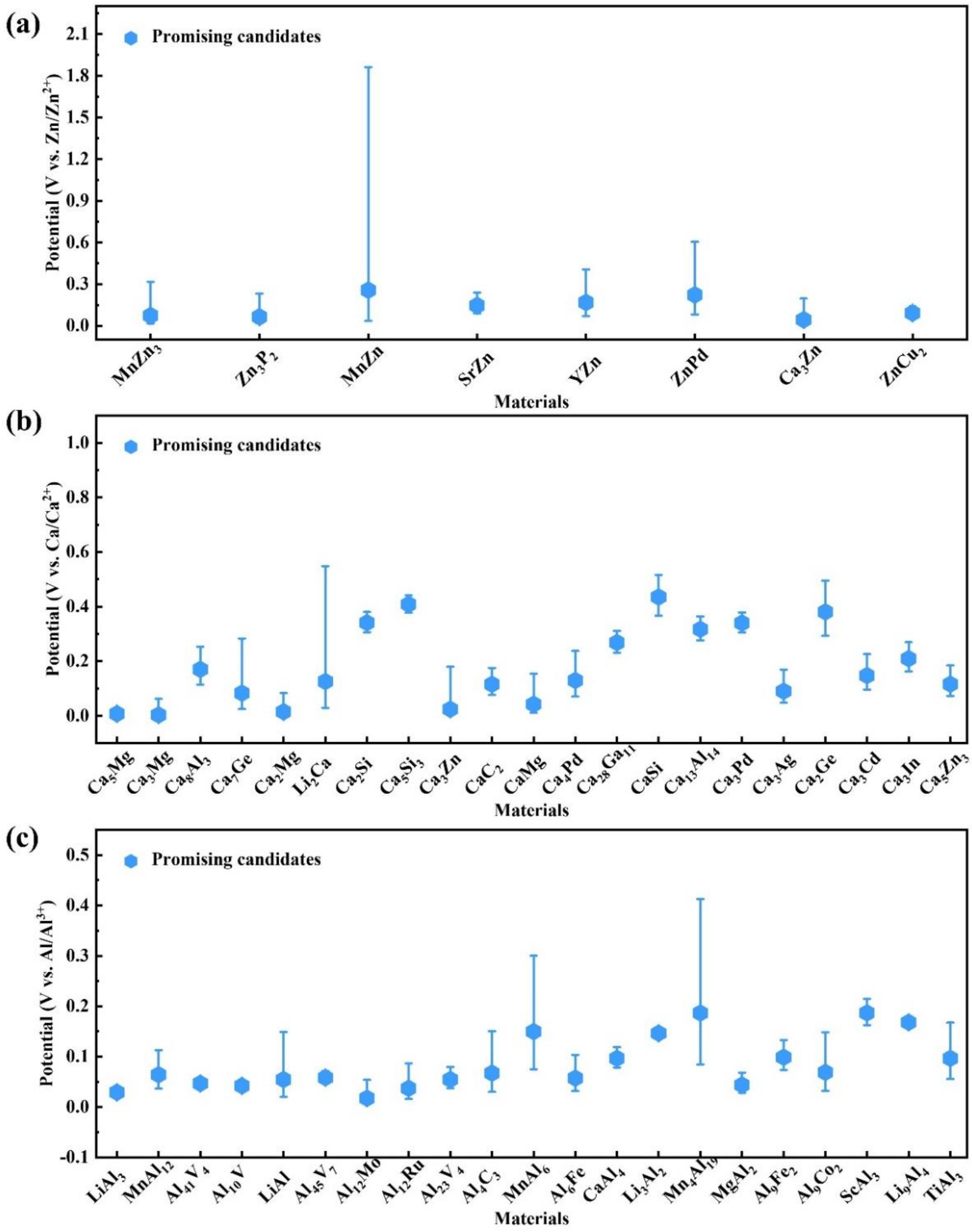

**Figure 5.** Candidate alloy anodes for Zn, Ca, and Al battery and their theoretical potentials.

In **Figure 5**, we present the theoretically calculated potentials of some Zn, Ca, and Al-based alloy anodes. We have identified candidate anodes with potentials controlled within the 0-0.5V range and organized them in descending order of specific capacity, including $MnZn_3$、$Zn_3P_2$、MnZn、$Ca_8Al_3$、$LiAl_3$、$MnAl_{12}$、$Al_{41}V_4$, etc. With the continuous development and advancement of experimental methods and technologies, these promising alloy anodes are expected to be practically applied in the future. Detailed calculated potential data can be found in the **Supporting Information**.

## Conclusions

Our study introduces an innovative approach that uses machine learning technology based on CGCNN to create an efficient screening tool for a variety of metal batteries. By combining potential data from MP and AFLOW databases, the tool accurately predicts and optimizes over 12,000 alloy anode materials, selecting hundreds with low potential and high capacity. The predicted performance of these materials aligns well with experimental data, providing a strong foundation for accuracy. We provide a detailed list of candidate materials for researchers and emphasize the need for further development to promote technological progress in related fields.

# Supplementary Information: Machine learning assisted screening of metal binary alloys for anode materials

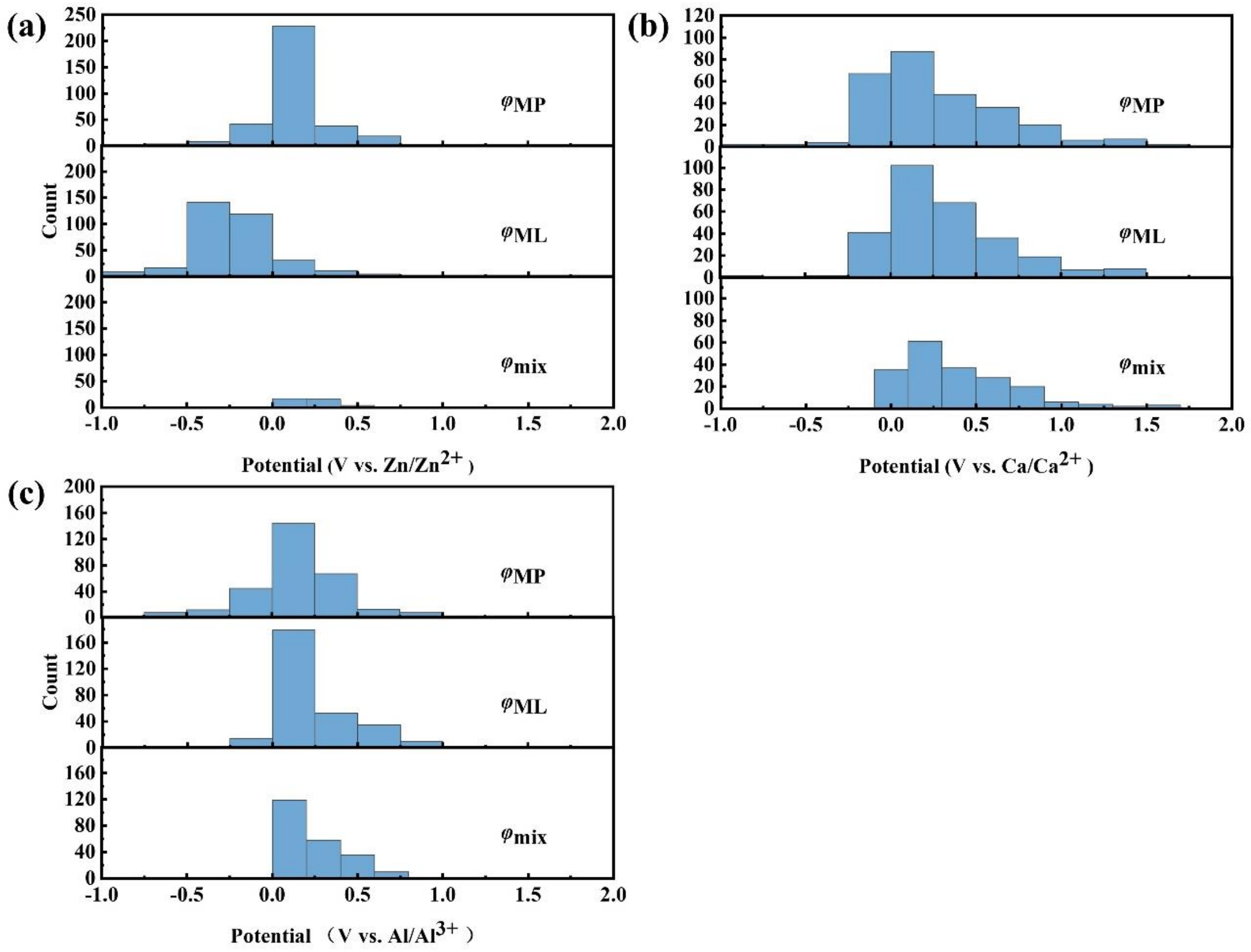

**Figure S1.** Calculated potentials and the quantitative distribution of the potentials. (a)(b)(c) are all made up of three small figures, from top to bottom: Statistical distributions of potentials calculated from energies from the MP database ( $\varphi_{MP}$ ); Potentials calculated from energies predicted by CGCNN-ML ($\varphi_{ML}$); Statistical distribution of $\varphi_{mix}$ ($\varphi_{mix} = \sqrt{\varphi_{MP}\ \varphi_{ML}}$).

**Table S1. $\varphi_{mix}$、 $\varphi_{experiment}$ and capacity of anode materials for Li batteries**

| Materials | $\varphi_{mix}$ （V） | $\varphi_{experiment}$ （V） | Capacity(mA h /g) |
|---|---|---|---|
| $Li_5Mg$ | 0.06683 | —— | 2269.869514 |
| $Li_4Mg$ | 0.05279 | —— | 2057.961551 |
| $Li_{22}Si_5$[1] | 0.21876 | 0.22 | 2010.554972 |
| $Li_{21}Si_5$ | 0.21290 | 0.36 | 1965.71171 |
| $Li_{15}Si_4$ | 0.21137 | 0.40 | 1856.402367 |
| $Li_7Si_2$ | 0.25668 | —— | 1790.041906 |
| $Li_3Mg$[2] | 0.07600 | 0.03 | 1780.867754 |
| $Li_{13}Si_4$ | 0.26985 | —— | 1719.133959 |
| $Li_3Al$ | 0.05731 | —— | 1681.14214 |
| $Li_9Al_4$ | 0.16025 | —— | 1414.936883 |
| LiC | 0.31081 | —— | 1413.518362 |
| $Li_7Si_3$ | 0.32027 | —— | 1411.592457 |
| $Li_2Mg$ | 0.12917 | —— | 1403.04292 |
| $Li_{12}Si_7$ | 0.43640 | —— | 1148.534802 |
| $Li_{22}Ge_5$ | 0.30546 | —— | 1142.383631 |
| $Li_3Al_2$ | 0.21865 | —— | 1074.611898 |
| $Li_{15}Ge_4$[3] | 0.33210 | 0.48 | 1018.141509 |
| $Li_2Ca$ | 0.11834 | —— | 992.920682 |
| $Li_7Ge_2$ | 0.37420 | —— | 967.2765349 |
| $Li_3Zn$ | 0.02199 | —— | 931.9858057 |
| $Li_3Ga$ | 0.26279 | —— | 887.5820025 |
| $Li_3Ge$ | 0.39949 | —— | 859.8803805 |
| LiMg | 0.15938 | —— | 857.3577418 |
| $Li_{13}In_3$ | 0.14765 | 0.30 | 801.1672767 |
| $Li_{22}Sn_5$[4] | 0.46525 | 0.50 | 789.7520559 |
| LiAl[5] | 0.29028 | 0.28 | 789.7002034 |
| $Li_{17}Sn_4$ | 0.47680 | 0.36 | 768.1874229 |
| LiSi | 0.10673 | —— | 764.8100037 |
| $Li_{15}Pd_4$ | 0.28743 | —— | 758.4726168 |
| $Li_2Ga$[6] | 0.42939 | 0.52 | 640.8468393 |
| $Li_3Pd$ | 0.35001 | —— | 631.6025243 |
| $Li_3Ag$ | 0.08372 | —— | 624.4958855 |
| $Li_3Cd$ | 0.12603 | —— | 603.2018854 |
| $Sr_{19}Li_{44}$ | 0.05048 | —— | 598.2771152 |
| $Li_3In$ | 0.23269 | —— | 592.4978436 |
| $Li_3Mg_5$ | 0.14801 | —— | 564.5811673 |
| $Li_3Sb$[7] | 0.82779 | 0.95 | 563.6506456 |
| $Li_9Ag_4$ | 0.16142 | —— | 488.1170018 |
| $LiMg_2$ | 0.15943 | —— | 482.2415438 |
| $Li_2In$ | 0.34635 | —— | 416.3014763 |
| LiZn[8] | 0.34571 | 0.38 | 370.2695232 |
| $Li_8Ag_5$ | 0.17324 | —— | 360.2681603 |
| $Li_3Bi$[7] | 0.62820 | 0.78 | 349.721283 |
| $Li_3Ag_2$ | 0.15941 | —— | 339.733428 |
| $LiMg_3$ | 0.12007 | —— | 335.4663394 |
| $Li_3In_2$ | 0.40879 | 0.12 | 320.8788664 |
| $Li_5In_4$[9] | 0.45021 | 0.34 | 271.1563494 |
| LiAg | 0.29214 | —— | 233.3353657 |
| LiCd | 0.48930 | —— | 224.4537167 |
| LiIn | 0.46275 | 0.62 | 220.0165901 |
| $Sr_3Li_2$ | 0.03283 | —— | 193.6027058 |
| $LiGa_2$ | 0.36673 | —— | 183.0012228 |
| $LiP_5$ | 0.36318 | —— | 165.5573478 |
| $LiCu_3$ | 0.25046 | —— | 135.5862718 |
| $LiZn_3$ | 0.16702 | —— | 131.8563947 |
| $LiSn_3$ | 0.44193 | —— | 73.78385673 |

**Table S2. Candidate anodes for Li batteries**

| **Candidate Materials** | **Predicted voltage (V)** | **Capacity (mA h $g^{-1}$)** |
|---|---|---|
| $Li_5Mg$ | 0.06730 | 2269.869514 |
| $Li_4Mg$ | 0.05279 | 2057.961551 |
| $Li_7Si_2$ | 0.25668 | 1790.041906 |
| $Li_{13}Si_4$ | 0.26985 | 1719.133959 |
| $Li_3Al$ | 0.05731 | 1681.14214 |
| $Li_9Al_4$ | 0.16025 | 1414.936883 |
| LiC | 0.31081 | 1413.518362 |
| $Li_7Si_3$ | 0.32027 | 1411.592457 |
| $Li_2Mg$ | 0.12981 | 1403.04292 |
| $Li_{12}Si_7$ | 0.43640 | 1148.534802 |
| $Li_{22}Ge_5$ | 0.30546 | 1142.383631 |
| $Li_3Al_2$ | 0.21865 | 1074.611898 |
| $Li_2Ca$ | 0.11843 | 992.920682 |
| $Li_7Ge_2$ | 0.37420 | 967.2765349 |
| $Li_3Zn$ | 0.02199 | 931.9858057 |
| $Li_3Ga$ | 0.26288 | 887.5820025 |
| $Li_3Ge$ | 0.39949 | 859.8803805 |
| LiMg | 0.16850 | 857.3577418 |
| LiSi | 0.17655 | 764.8100037 |
| $Li_{15}Pd_4$ | 0.28743 | 758.4726168 |
| $Li_3Pd$ | 0.35069 | 631.6025243 |
| $Li_3Ag$ | 0.08462 | 624.4958855 |
| $Li_3Cd$ | 0.13028 | 603.2018854 |
| $Sr_{19}Li_{44}$ | 0.05048 | 598.2771152 |

**Table S3. $\varphi_{mix}$、$\varphi_{experiment}$ and capacity of anode materials for Na batteries**

| Materials | $\varphi_{mix}$ (V) | $\varphi_{experiment}$ (V) | Capacity (mAh $g^{-1}$) |
|---|---|---|---|
| $Na_3Mg$ | 0.051001834 | —— | 861.6135084 |
| $Na_3Ca$ | 0.07187449 | —— | 736.9873817 |
| $KNa_2$ | 0.234330541 | —— | 629.7515221 |
| $Na_3Zn$ | 0.070421391 | —— | 598.0621972 |
| $Na_3Ga$ | 0.12588788 | —— | 579.4596699 |
| $Na_3Ge$ | 0.278272179 | —— | 567.52348 |
| $RbNa_3$ | 0.106809095 | —— | 520.3835844 |
| $Na_{15}Sn_4$[10] | 0.323778947 | 0.3 | 490.2256208 |
| $Na_3Pd$ | 0.148568647 | —— | 458.2188266 |
| $Na_3Ag$ | 0.098364256 | —— | 454.4668001 |
| $Na_3Cd$ | 0.186605355 | —— | 443.0838952 |
| $Na_3In$ | 0.186830868 | —— | 437.2809977 |
| $Na_2Se$[11] | 1.749635077 | 1.85 | 428.8298383 |
| $Na_3Sn$ | 0.357856554 | —— | 428.210634 |
| $Na_3Sb$[12] | 0.695178127 | 0.57 | 421.3652808 |
| $Na_5Sn_2$ | 0.458319703 | —— | 380.1238464 |
| $Na_9Sn_4$[10] | 0.487932286 | 0.4 | 353.6480901 |
| $Na_2In$[13] | 0.262321602 | 0.31 | 333.2006617 |
| $Na_2Sn$ | 0.353512104 | —— | 325.3243954 |
| $BaNa_2$ | 0.17708497 | —— | 292.2856192 |
| $Na_3Bi$[14] | 0.561572237 | 0.41 | 289.141932 |
| NaGa | 0.347050165 | —— | 288.9454553 |
| NaGe[15] | 0.796856061 | 0.19 | 280.1317578 |
| NaIn | 0.363265344 | —— | 194.3936491 |
| $K_3Na$ | 0.330650295 | —— | 190.9626187 |
| BaNa | 0.222970039 | —— | 167.1001828 |
| NaBi[14] | 0.660374034 | 0.67 | 115.484761 |
| $NaGa_3$ | 0.177507456 | —— | 115.3907451 |
| $NaIn_2$ | 0.016975119 | —— | 106.0421334 |
| $Rb_3Na$ | 0.03298506 | —— | 95.88251716 |
| $Na_5Si_{51}$ | 0.361693871 | —— | 86.56490898 |
| $NaPd_3$ | 0.358826899 | —— | 78.27319211 |
| $NaAg_3$ | 0.302368582 | —— | 77.29216316 |
| $NaCd_3$ | 0.147595974 | —— | 74.36782215 |
| $NaZn_{13}$ | 0.248666273 | —— | 30.67535242 |

**Table S4. Candidate anodes for Na batteries**

| Candidate Materials | Predicted $\varphi_{mix}$ (V) | Capacity (mA h/g) |
|---|---|---|
| $Na_3Mg$ | 0.051001834 | 861.6135084 |
| $Na_3Ca$ | 0.07187449 | 736.9873817 |
| $KNa_2$ | 0.234330541 | 629.7515221 |
| $Na_3Zn$ | 0.070421391 | 598.0621972 |
| $Na_3Ga$ | 0.12588788 | 579.4596699 |
| $Na_3Ge$ | 0.278272179 | 567.52348 |
| $RbNa_3$ | 0.106809095 | 520.3835844 |
| $Na_3Pd$ | 0.148568647 | 458.2188266 |
| $Na_3Ag$ | 0.098364256 | 454.4668001 |
| $Na_3Cd$ | 0.186605355 | 443.0838952 |
| $Na_3In$ | 0.186830868 | 437.2809977 |
| $Na_3Sn$ | 0.357856554 | 428.210634 |
| $Na_5Sn_2$ | 0.458319703 | 380.1238464 |
| $Na_2Sn$ | 0.353512104 | 325.3243954 |
| $BaNa_2$ | 0.17708497 | 292.2856192 |
| NaGa | 0.347050165 | 288.9454553 |

**Table S5. $\varphi_{mix}$、$\varphi_{experiment}$ and capacity of anode materials for K batteries**

| Materials | $\varphi_{mix}$ (V) | $\varphi_{experiment}$ (V) | Capacity (mAh $g^{-1}$) |
|---|---|---|---|
| $K_3Na$ | 0.046090014 | —— | 572.8878561 |
| $K_3P$ | 0.552284643 | —— | 542.0387407 |
| $K_4P_3$[16] | 0.975719933 | 1.00 | 429.8033805 |
| KSi | 0.383235333 | —— | 398.7407716 |
| $K_3Rb$ | 0.025614192 | —— | 396.3612511 |
| KP | 1.161622485 | —— | 382.3067702 |
| $K_3Cd$ | 0.125076826 | —— | 349.8704861 |
| $K_3In$ | 0.054475909 | —— | 346.2423313 |
| $K_2Se$ | 1.554832472 | —— | 340.9223956 |
| $K_3Sn$ | 0.282748335 | —— | 340.530921 |
| $K_3Sb$[17] | 0.729943552 | 0.69 | 336.1876396 |
| $KNa_2$ | 0.409828261 | —— | 314.8757611 |
| $K_2P_3$ | 1.370683496 | —— | 313.1055763 |
| $K_{57}Se_{34}$ | 1.738129088 | —— | 310.7882682 |
| $K_5Se_3$ | 1.756004217 | —— | 309.7925388 |
| $K_2Te$ | 1.359135748 | —— | 260.3451962 |
| $K_3Bi$[18] | 0.445960297 | 0.40 | 246.3175123 |
| KGa | 0.376798035 | —— | 246.1749111 |
| KGe | 0.663641877 | —— | 239.7483399 |
| $K_5Te_3$ | 1.616968211 | —— | 231.6225423 |
| KSe | 1.3982231 | —— | 226.9138898 |
| $KC_8$ | 2.131883441 | —— | 198.1640111 |
| $K_5Sb_4$ | 1.054537539 | —— | 196.247784 |
| $K_2Ga_3$ | 0.746297052 | —— | 186.4458093 |
| $K_3P_{11}$ | 1.463327522 | —— | 175.4707341 |
| $K_2Se_3$ | 2.259401271 | —— | 170.0478615 |
| KSn[19] | 0.758567819 | 0.88 | 169.7558441 |
| KSb[20] | 1.269945095 | 1.12 | 166.538189 |
| KTe | 1.715233555 | —— | 160.7037877 |
| $K_3Bi_2$ | 0.686209342 | —— | 150.1474067 |
| $K_8In_{11}$ | 0.487337224 | —— | 136.0035843 |
| $K_4Si_{23}$ | 0.881139544 | —— | 133.54936 |
| $K_5Bi_4$ | 0.737432583 | —— | 129.8659117 |
| $K_{15}Si_{92}$ | 0.822091899 | —— | 126.7465561 |
| $K_2Te_3$ | 2.043309344 | —— | 116.2222666 |
| $K_2Se_5$ | 2.407262142 | —— | 113.2736852 |
| KBi | 0.808179846 | —— | 107.9861979 |
| $KGa_3$ | 1.15714477 | —— | 107.9039905 |
| $KAg_2$ | 0.363178698 | —— | 105.1233352 |
| $KSe_3$ | 2.399527832 | —— | 97.06933161 |
| $KSn_2$ | 0.747307486 | —— | 96.87906842 |
| $KSb_2$[20] | 1.5354376 | 1.12 | 94.78872542 |
| $K_4Sn_9$ | 0.919507142 | —— | 87.48921245 |
| $K_{17}In_{41}$ | 0.538366963 | —— | 84.77209726 |
| $K_3Ga_{13}$ | 1.483834715 | —— | 78.50693518 |
| $KCd_3$ | 0.41991495 | —— | 71.18467519 |
| $KIn_3$ | 0.475846471 | —— | 69.84450609 |
| $KSn_3$ | 0.729434025 | —— | 67.78061437 |
| $KSb_3$ | 0.23086295 | —— | 66.24742197 |
| $KTe_3$ | 1.272391791 | —— | 63.49639012 |
| $K_3Ge_{17}$ | 1.772819935 | —— | 59.43539811 |
| $K_4Ge_{23}$ | 1.70686664 | —— | 58.64774573 |
| $kBi_2$ | 1.029302145 | —— | 58.61181732 |
| $KIn_4$ | 0.375792159 | —— | 53.75323555 |
| $K_6Sn_{25}$ | 0.953752859 | —— | 50.19231112 |
| $K_3Cd_{16}$ | 1.295601668 | —— | 41.94804449 |
| $KBi_3$ | 0.504217159 | —— | 40.22142881 |
| $K_4Sn_{23}$ | 0.984001572 | —— | 37.12000887 |
| $KCd_{13}$ | 1.869582721 | —— | 17.85408423 |

**Table S6. Candidate anodes for K batteries**

| Candidate Materials | Predicted $\varphi_{mix}$ (V) | Capacity (mA h $g^{-1}$) |
|---|---|---|
| $K_3Na$ | 0.046090014 | 572.8878561 |
| $K_3P$ | 0.552284643 | 542.0387407 |
| KSi | 0.383235333 | 398.7407716 |
| $K_3Rb$ | 0.025614192 | 396.3612511 |
| $K_3Cd$ | 0.125076826 | 349.8704861 |
| $K_3In$ | 0.054475909 | 346.2423313 |
| $K_3Sn$ | 0.282748335 | 340.530921 |
| $KNa_2$ | 0.409828261 | 314.8757611 |
| KGa | 0.376798035 | 246.1749111 |
| $K_8In_{11}$ | 0.487337224 | 136.0035843 |
| $KAg_2$ | 0.363178698 | 105.1233352 |

**Table S7. $\varphi_{mix}$、$\varphi_{experiment}$ and capacity of anode materials for Zn batteries**

| Materials | $\varphi_{mix}$ (V) | $\varphi_{experiment}$ (V) | Capacity (mAh $g^{-1}$) |
|---|---|---|---|
| $MnZn_3$ | 0.073667812 | —— | 639.9538152 |
| $Zn_3P_2$ | 0.064816485 | —— | 622.5777089 |
| MnZn | 0.256258386 | —— | 445.195975 |
| SrZn | 0.146097487 | —— | 350.1166446 |
| YZn | 0.168445785 | —— | 344.9634614 |
| ZnPd | 0.222374672 | —— | 311.8099971 |
| $Ca_3Zn$ | 0.043354888 | —— | 288.607704 |
| $ZnCu_2$ | 0.091804476 | —— | 278.3258269 |
| ZnTe[21] | 0.175377135 | 0.23 | 277.5932729 |
| BaZn | 0.16224378 | —— | 264.2747218 |
| $ZnFe_3$ | 0.298917759 | —— | 230.0037777 |
| $Zr_2Zn$ | 0.380389476 | —— | 216.1649661 |
| $ZnCu_3$ | 0.111292972 | —— | 209.2506454 |
| $ZnPd_2$ | 0.30267628 | —— | 192.5541511 |
| $ZnNi_4$ | 0.21568421 | —— | 178.485647 |
| $ZnTe_2$ | 0.199783637 | —— | 167.1132127 |
| $Zr_3Zn$ | 0.340975984 | —— | 158.0094432 |
| $Ba_2Zn$ | 0.270177656 | —— | 157.5531593 |
| $ZnPd_3$ | 0.373944886 | —— | 139.2833839 |
| $ZnPd_4$ | 0.229110507 | —— | 109.1003871 |

**Table S8. Candidate anodes for Zn batteries**

| **Candidate Materials** | **Predicted $\varphi_{mix}$ (V)** | **Capacity (mA h $g^{-1}$)** |
|---|---|---|
| $MnZn_3$ | 0.073667812 | 639.9538152 |
| $Zn_3P_2$ | 0.064816485 | 622.5777089 |
| MnZn | 0.256258386 | 445.195975 |
| SrZn | 0.146097487 | 350.1166446 |
| YZn | 0.168445785 | 344.9634614 |
| ZnPd | 0.222374672 | 311.8099971 |
| $Ca_3Zn$ | 0.043354888 | 288.607704 |
| $ZnCu_2$ | 0.091804476 | 278.3258269 |

**Table S9. $\varphi_{mix}$、$\varphi_{experiment}$ and capacity of anode materials for Mg batteries**

| **Materials** | **$\varphi_{mix}$ (V)** | **$\varphi_{experiment}$ (V)** | **Capacity (mAh $g^{-1}$)** |
|---|---|---|---|
| $Mg_{149}Mn$ | 0.006670927 | —— | 2171.460917 |
| $BaMg_{149}$ | 0.005384508 | —— | 2123.86439 |
| $LiMg_5$ | 0.011540623 | —— | 2085.298834 |
| $LiMg_3$ | 0.017825219 | —— | 2012.798036 |
| $CaMg_{15}$ | 0.005021952 | —— | 1986.072017 |
| $LiMg_2$ | 0.046570821 | —— | 1928.966175 |
| $Li_3Mg_5$ | 0.059052864 | —— | 1881.937224 |
| $CaMg_7$ | 0.027959392 | —— | 1784.123722 |
| $YMg_{15}$ | 0.028192679 | —— | 1768.324748 |
| $ZrMg_{15}$ | 0.006651725 | —— | 1763.211416 |
| $ScMg_7$ | 0.031233669 | —— | 1743.661985 |
| LiMg | 0.0745185 | —— | 1714.715484 |
| $SnMg_{15}$ | 0.008779638 | —— | 1662.92837 |
| $CaMg_5$ | 0.022970059 | —— | 1657.704374 |
| $SbMg_{15}$ | 0.037218526 | —— | 1652.502904 |
| $ScMg_5$ | 0.036208612 | —— | 1609.132574 |
| $LaMg_{15}$ | 0.016131765 | —— | 1596.227067 |
| $Sr_2Mg_{17}$ | 0.045564954 | —— | 1547.905001 |
| $ScMg_4$ | 0.038679101 | —— | 1507.371146 |
| $YMg_7$ | 0.033280017 | —— | 1442.257183 |
| $Y_{11}Mg_{76}$ | 0.05990049 | —— | 1435.725805 |
| $ZnMg_5$ | 0.016491657 | —— | 1433.072635 |
| $CaMg_3$ | 0.026566669 | —— | 1422.512899 |
| $Li_2Mg$ | 0.103682123 | —— | 1403.04292 |
| $BiMg_{15}$ | 0.020611924 | —— | 1401.208254 |
| $SiMg_2$ | 0.009396825 | —— | 1397.152394 |
| $Y_4Mg_{25}$ | 0.06130275 | —— | 1384.803706 |
| $Mg_5Ge$ | 0.009638038 | —— | 1379.70283 |
| $Mg_{87}Pb_{12}$ | 0.076738376 | —— | 1013.117116 |
| $Mg_3Sc$ | 0.092472519 | —— | 1363.643305 |
| $Mg_2P$ | 0.09517429 | —— | 1346.451548 |
| $Mg_9Si_5$ | 0.003607783 | —— | 1342.526624 |
| $Mg_5Se$ | 0.258374213 | —— | 1336.209691 |
| $Mg_7Si_4$ | 0.008653995 | —— | 1327.695156 |
| $Mg_7C_d$ | 0.014958707 | —— | 1327.38032 |
| $Ba_2Mg_{17}$ | 0.047781416 | —— | 1324.184875 |
| $Mg_{18}Al_{11}$ | 0.033720568 | —— | 1313.379419 |
| $Mg_7Sn$ | 0.043087566 | —— | 1298.428921 |
| $Mg_7Sb$ | 0.097385109 | —— | 1284.866133 |
| $SrMg_5$ | 0.047253283 | —— | 1280.881685 |
| $Y_{19}Mg_{97}$ | 0.069555904 | —— | 1278.239775 |
| $YMg_5$ | 0.044045759 | —— | 1267.032744 |
| $Si_2Mg_3$ | 0.004249276 | —— | 1245.160241 |
| $Y_5Mg_{24}$ | 0.075073194 | —— | 1244.974585 |
| $LaMg_7$ | 0.018536893 | —— | 1213.580075 |
| $CaMg_2$ | 0.0345625 | —— | 1208.235612 |
| $P_2Mg_3$ | 0.45060768 | —— | 1191.831711 |
| $Sr_9Mg_{38}$ | 0.073630401 | —— | 1189.113231 |
| $Li_3Mg$ | 0.147353653 | —— | 1187.245169 |
| $Mg_4Si_3$ | 0.002970234 | —— | 1180.925512 |
| $Mg_5Pd$ | 0.07290654 | —— | 1175.239641 |
| $Mg_3Zn$ | 0.016480097 | —— | 1162.010931 |
| $SrMg_4$ | 0.064658257 | —— | 1159.446007 |
| $Al_{13}Mg_{16}$ | 0.017223861 | —— | 1158.997683 |
| $Mg_2Sc$ | 0.100800679 | —— | 1145.245068 |
| $Mg_5Cd$ | 0.034507308 | —— | 1145.142261 |
| $Sr_6Mg_{23}$ | 0.080122697 | —— | 1136.032303 |
| $Mg_5In$ | 0.049323671 | —— | 1133.479731 |
| $Mg_3Ga$ | 0.051387492 | —— | 1126.866613 |
| $Mg_6Si_5$ | 0.004390538 | —— | 1122.993083 |
| $Mg_5Sn$ | 0.059598699 | —— | 1115.111807 |
| $Mg_3Ge$ | 0.068912422 | —— | 1104.283604 |
| $Mg_5Sb$ | 0.115208134 | —— | 1101.136527 |
| $Mg_5Te$ | 0.157649508 | —— | 1075.323633 |
| MgAl | 0.020259062 | —— | 1044.670189 |
| $Li_4Mg$ | 0.128550814 | —— | 1028.980776 |
| $LaMg_5$ | 0.028600411 | —— | 1028.640984 |
| $Mg_2Fe$ | 0.054722777 | —— | 1025.858025 |
| MgSi | 0.005376071 | —— | 1022.656563 |
| $Mg_{26}Ag_7$ | 0.049205166 | —— | 1004.341726 |
| $SrMg_3$ | 0.02006267 | —— | 1001.239605 |
| $Mg_2Ni$ | 0.031540972 | —— | 998.6300476 |
| $Mg_7Bi$ | 0.055388352 | —— | 989.2671089 |
| $YMg_3$ | 0.026893073 | —— | 987.1822431 |
| $Mg_3Zr$ | 0.013416522 | —— | 979.2553872 |
| $Mg_2Zn$ | 0.010678674 | —— | 939.8082776 |
| $Mg_5Si_6$ | 0.010182437 | —— | 923.6280388 |
| $Mg_{54}Ag_{17}$ | 0.074851419 | —— | 919.5817465 |
| $Li_5Mg$ | 0.194349586 | —— | 907.9478055 |
| $Mg_2Ga$ | 0.019367661 | —— | 905.5462128 |
| $Mg_{23}Al_{30}$ | 0.053453076 | —— | 900.4870385 |
| $Mg_3Pd$ | 0.213554827 | —— | 896.2779156 |
| $Ba_6Mg_{23}$ | 0.083618959 | —— | 891.0444642 |
| $Mg_2Ge$ | 0.095022588 | —— | 883.7608247 |
| $Mg_3Si_4$ | 0.003377637 | —— | 867.6177676 |
| $Mg_3Cd$ | 0.055179834 | —— | 867.304102 |
| $Mg_3In$ | 0.067801689 | —— | 856.1840486 |
| $Mg_3Sn$ | 0.098271762 | —— | 838.7901433 |
| CaMg | 0.043029098 | —— | 832.1761956 |
| $Mg_3Sb$ | 0.12462906 | —— | 825.6530114 |
| $Mg_5Bi$ | 0.045055065 | —— | 810.5474955 |
| $Mg_2Si_3$ | 0.013425901 | —— | 806.4846314 |
| $Mg_3Te$ | 0.259220035 | —— | 801.6058649 |
| $Mg_5Pd_2$ | 0.219020529 | —— | 801.190316 |
| $SrMg_2$ | 0.030909454 | —— | 786.5815166 |
| $YMg_2$ | 0.022541818 | —— | 773.6001617 |
| MgSc | 0.187799899 | —— | 773.5666537 |
| $BaMg_3$ | 0.028759736 | —— | 764.5189829 |
| $Mg_5In_2$ | 0.095989648 | —— | 762.869453 |
| $LaMg_3$ | 0.061735749 | —— | 758.819947 |
| $Mg_4Si_7$ | 0.033668006 | —— | 729.3939868 |
| $Mg_2Pd$ | 0.295163869 | —— | 691.1952525 |
| $Mg_2Ag$ | 0.080106324 | —— | 684.7991411 |
| $MgAl_2$ | 0.112387801 | —— | 684.536662 |
| $MgSi_2$ | 0.011859243 | —— | 665.7554332 |
| $Mg_2Cd$ | 0.061952719 | —— | 665.4784159 |
| $Mg_2In$ | 0.114538656 | —— | 655.6771178 |
| MgNi | 0.030112561 | —— | 645.5336273 |
| $Mg_2Sn$[22] | 0.083551112 | 0.15 | 640.4217044 |
| $Mg_2Sb$ | 0.16899275 | —— | 628.9604977 |
| MgCu | 0.132405237 | —— | 609.8735359 |
| MgZn | 0.043348551 | —— | 597.2089083 |
| $Mg_9Sn_5$ | 0.182201981 | —— | 593.6255078 |
| $BaMg_2$ | 0.098328973 | —— | 576.3027262 |
| $Na_3Mg$ | 0.087025253 | —— | 551.1005966 |
| $LaMg_2$ | 0.008397399 | —— | 571.4499029 |
| $Mg_3Bi$ | 0.048632875 | —— | 570.1910286 |
| MgGa | 0.009359638 | —— | 569.8089931 |
| MgGe | 0.093712555 | —— | 552.6638816 |
| $Mg_{21}Zn_{25}$ | 0.103888465 | —— | 524.3858447 |
| $Ca_2Mg$ | 0.028776966 | —— | 512.899551 |
| $Mg_3Sb_2$[23] | 0.433978007 | 0.32 | 507.9526601 |
| SrMg | 0.037250213 | —— | 478.6955551 |
| YMg | 0.046328271 | —— | 469.1141834 |
| $MgSc_2$ | 0.299985787 | —— | 469.0895401 |
| MgZr | 0.030216204 | —— | 463.7623454 |
| $Mg_2Bi$ | 0.081690589 | —— | 415.9944097 |
| MgAg | 0.151456785 | —— | 405.3626686 |
| MgCd | 0.094682277 | —— | 391.8926826 |
| $Mg_4Zn_7$ | 0.161542824 | —— | 386.0900082 |
| MgIn[24] | 0.15736015 | 0.09 | 385.1124544 |
| $MgNi_2$ | 0.143997842 | —— | 378.1326972 |
| MgSn | 0.117697695 | —— | 374.6294121 |
| $Ca_3Mg$ | 0.008173119 | —— | 370.6819613 |
| MgSb | 0.28218672 | —— | 366.8092972 |
| $MgCu_2$ | 0.1744668 | —— | 353.890764 |
| $MgZn_2$ | 0.076426052 | —— | 345.3904321 |
| $MgSc_3$ | 0.327345798 | —— | 336.6023132 |
| BaMg | 0.108580856 | —— | 331.4813898 |
| LaMg | 0.010124741 | —— | 328.2744423 |
| $Mg_3Bi_2$[22-23, 25] | 0.236183416 | 0.23 | 327.4438503 |
| $MgGe_2$ | 0.151992721 | —— | 315.9359613 |
| $Mg_2Ga_5$[26] | 0.287713893 | 0.3 | 269.7614702 |
| $Sr_2Mg$ | 0.022123277 | —— | 268.5008394 |
| $MgNi_3$ | 0.26066336 | —— | 267.3766369 |
| $Y_2Mg$ | 0.127881443 | —— | 262.4867111 |
| $MgZr_2$ | 0.062588646 | —— | 259.1401334 |
| $MgCu_3$ | 0.25577571 | —— | 249.2660845 |
| $MgZn_3$ | 0.117862884 | —— | 242.9488691 |
| $Ca_5Mg$ | 0.05081904 | —— | 238.4476735 |
| MgBi | 0.09267177 | —— | 229.667574 |
| $MgGa_3$ | 0.08956856 | —— | 229.4816553 |
| $MgAg_2$ | 0.177968197 | —— | 223.2035361 |
| $MgGe_3$ | 0.014142199 | —— | 221.1910414 |
| $MgCd_2$ | 0.141174557 | —— | 215.063 |
| $MgIn_2$ | 0.160032171 | —— | 210.9860164 |
| $MgSn_2$ | 0.195236414 | —— | 204.7094874 |
| $MgSb_2$ | 0.359108896 | —— | 200.0485392 |
| $Y_3Mg$ | 0.065900999 | —— | 182.2238396 |
| $MgZr_3$ | 0.268297338 | —— | 179.8058239 |
| $La_2Mg$ | 0.030262767 | —— | 177.3418907 |
| $MgNi_5$ | 0.316358384 | —— | 168.6062246 |
| $Mg_2Bi_3$ | 0.183078042 | —— | 158.6203834 |
| $MgCu_5$ | 0.285142586 | —— | 156.6447878 |
| $MgPd_3$ | 0.427568698 | —— | 155.9472007 |
| $MgAg_3$ | 0.126408657 | —— | 154.0000402 |
| $MgZn_5$ | 0.143489052 | —— | 152.4918173 |
| $MgCd_3$ | 0.184702615 | —— | 148.1946573 |
| $MgIn_3$ | 0.198879638 | —— | 145.2927251 |
| $MgGa_5$ | 0.164842833 | —— | 143.6715649 |
| $MgSn_3$ | 0.150328362 | —— | 140.832409 |
| $Mg_2Zn_{11}$ | 0.274094084 | —— | 139.506242 |
| $MgSb_3$ | 0.46567686 | —— | 137.5258288 |
| $La_3Mg$ | 0.141146168 | —— | 121.4857275 |
| $MgBi_2$ | 0.01132859 | —— | 121.1445626 |
| $MgAg_4$ | 0.121004191 | —— | 117.5531016 |
| $Y_5Mg$ | 0.079386137 | —— | 113.0731162 |
| $MgAg_5$ | 0.275677701 | —— | 95.05628543 |
| $La_4Mg$ | 0.037495549 | —— | 92.38717153 |
| $MgCd_5$ | 0.239696989 | —— | 91.37389999 |
| $MgIn_5$ | 0.153852657 | —— | 89.53617594 |
| $MgSn_5$ | 0.215567548 | —— | 86.71543715 |

**Table S10. Candidate anodes for Mg batteries**

| Candidate Materials | Predicted $\varphi_{mix}$ (V) | Capacity (mA h $g^{-1}$) |
|---|---|---|
| $Mg_{149}Mn$ | 0.006670927 | 2171.460917 |
| $BaMg_{149}$ | 0.005384508 | 2123.86439 |
| $LiMg_5$ | 0.011540623 | 2085.298834 |
| $LiMg_3$ | 0.017825219 | 2012.798036 |
| $CaMg_{15}$ | 0.005021952 | 1986.072017 |
| $LiMg_2$ | 0.046570821 | 1928.966175 |
| $Li_3Mg_5$ | 0.059052864 | 1881.937224 |
| $CaMg_7$ | 0.027959392 | 1784.123722 |
| $YMg_{15}$ | 0.028192679 | 1768.324748 |
| $ZrMg_{15}$ | 0.006651725 | 1763.211416 |
| $ScMg_7$ | 0.031233669 | 1743.661985 |
| LiMg | 0.0745185 | 1714.715484 |
| $SnMg_{15}$ | 0.008779638 | 1662.92837 |
| $CaMg_5$ | 0.022970059 | 1657.704374 |
| $SbMg_{15}$ | 0.037218526 | 1652.502904 |
| $ScMg_5$ | 0.036208612 | 1609.132574 |
| $LaMg_{15}$ | 0.016131765 | 1596.227067 |
| $Sr_2Mg_{17}$ | 0.045564954 | 1547.905001 |
| $ScMg_4$ | 0.038679101 | 1507.371146 |
| $YMg_7$ | 0.033280017 | 1442.257183 |
| $ZnMg_5$ | 0.016491657 | 1433.072635 |

**Table S11. $\varphi_{mix}$、$\varphi_{experiment}$ and capacity of anode materials for Ca batteries**

| Materials | $\varphi_{mix}$ (V) | $\varphi_{experiment}$ (V) | Capacity (mAh $g^{-1}$) |
|---|---|---|---|
| $Ca_5Mg$ | 0.007519558 | —— | 1192.238368 |
| $Ca_3Mg$ | 0.002879026 | —— | 1112.045884 |
| $Ca_8Al_3$ | 0.169840692 | —— | 1067.370571 |
| $Ca_7Ge$ | 0.082960854 | —— | 1061.893733 |
| $Ca_2Mg$ | 0.014673173 | —— | 1025.799102 |
| $Li_2Ca$ | 0.125455308 | —— | 992.920682 |
| $Ca_2Si$ | 0.340936462 | —— | 989.966926 |
| $Ca_5Si_3$ | 0.408235676 | —— | 941.127287 |
| $Ca_3Zn$ | 0.024003199 | —— | 865.8231121 |
| $CaC_2$ | 0.115546942 | —— | 835.850234 |
| CaMg | 0.042131729 | —— | 832.1761956 |
| $Ca_4Pd$ | 0.129176962 | —— | 803.4731491 |
| $Ca_{28}Ga_{11}$ | 0.268015129 | —— | 794.11075 |
| CaSi | 0.435122119 | —— | 786.0160789 |
| $Ca_{13}Al_{14}$ | 0.317079036 | —— | 774.9704594 |
| $Ca_3Pd$ | 0.339963281 | —— | 709.1602178 |
| $Ca_3Ag$ | 0.090076797 | —— | 704.6584423 |
| $Ca_2Ge$ | 0.380518224 | —— | 701.3010812 |
| $Ca_3Cd$ | 0.147297862 | —— | 690.8981495 |
| $Ca_3In$ | 0.209212071 | —— | 683.823154 |
| $Ca_5Zn_3$ | 0.115732488 | —— | 675.4375128 |
| $Ca_3Sn$ | 0.352533187 | —— | 672.6819979 |
| $Ca_5Ga_3$ | 0.33575266 | —— | 654.0937936 |
| $Ca_5Pd_2$ | 0.400844913 | —— | 648.2830385 |
| $Ca_8In_3$ | 0.247357799 | —— | 644.4717762 |
| $Ca_5Ge_3$ | 0.480244394 | —— | 640.4102221 |
| $Ca_{11}Ga_7$ | 0.362381913 | —— | 634.4557491 |
| $CaMg_2$ | 0.066361932 | —— | 604.1178062 |
| $CaAl_2$ | 0.469841437 | —— | 569.7241658 |
| $Ca_2In$ | 0.301461882 | —— | 549.5912275 |
| CaNi | 0.018638906 | —— | 542.4466696 |
| $Ca_2Sn$ | 0.416569187 | —— | 538.832486 |
| CaCu | 0.15790546 | —— | 517.042384 |
| $Ca_5Ag_3$ | 0.131210643 | —— | 511.246312 |
| CaZn | 0.060279662 | —— | 507.9109274 |
| $Na_3Ca$ | 0.117935244 | —— | 491.3249211 |
| CaGa | 0.490664938 | —— | 487.9554831 |
| $CaMg_3$ | 0.075330483 | —— | 474.1709663 |
| $Ca_3Cd_2$ | 0.280783103 | —— | 465.8200408 |
| CaAg | 0.247591417 | —— | 362.1456477 |
| CaCd | 0.438811062 | —— | 351.3564913 |
| CaIn | 0.476847913 | —— | 345.8966016 |
| $CaNi_2$ | 0.312465928 | —— | 340.2555505 |
| $CaMg_5$ | 0.102450623 | —— | 331.5408749 |
| $CaZn_2$ | 0.273015845 | —— | 313.5123116 |
| $Ca_3Bi_2$ | 0.417433063 | —— | 298.6543886 |
| $CaMg_7$ | 0.165636396 | —— | 254.8748174 |
| $CaNi_3$ | 0.299990944 | —— | 247.8661343 |
| $CaZn_3$ | 0.225515162 | —— | 226.7324009 |
| $Ca_2Ni_7$ | 0.294342215 | —— | 218.2372145 |
| $CaAg_2$ | 0.490701932 | —— | 209.4412346 |
| $CaIn_2$ | 0.222150492 | —— | 198.6474562 |
| $Ca_3Ag_8$ | 0.44060991 | —— | 163.48413 |
| $CaNi_5$ | 0.032122903 | —— | 160.6329619 |
| $CaCu_5$ | 0.261875013 | —— | 149.7395251 |
| $CaAg_3$ | 0.198618955 | —— | 147.3210112 |
| $CaZn_5$ | 0.409389385 | —— | 145.940189 |
| $Ca_2Ag_7$ | 0.370587263 | —— | 128.2948929 |
| $CaZn_{13}$ | 0.325196775 | —— | 60.17329388 |

**Table S12. Candidate anodes for Ca batteries**

| Candidate Materials | Predicted $\varphi_{mix}$ (V) | Capacity (mA h $g^{-1}$) |
|---|---|---|
| $Ca_5Mg$ | 0.007519558 | 1192.238368 |
| $Ca_3Mg$ | 0.002879026 | 1112.045884 |
| $Ca_8Al_3$ | 0.169840692 | 1067.370571 |
| $Ca_7Ge$ | 0.082960854 | 1061.893733 |
| $Ca_2Mg$ | 0.014673173 | 1025.799102 |
| $Li_2Ca$ | 0.125455308 | 992.920682 |
| $Ca_2Si$ | 0.340936462 | 989.966926 |
| $Ca_5Si_3$ | 0.408235676 | 941.127287 |
| $Ca_3Zn$ | 0.024003199 | 865.8231121 |
| $CaC_2$ | 0.115546942 | 835.850234 |
| CaMg | 0.042131729 | 832.1761956 |
| $Ca_4Pd$ | 0.129176962 | 803.4731491 |
| $Ca_{28}Ga_{11}$ | 0.268015129 | 794.11075 |
| CaSi | 0.435122119 | 786.0160789 |
| $Ca_{13}Al_{14}$ | 0.317079036 | 774.9704594 |
| $Ca_3Pd$ | 0.339963281 | 709.1602178 |
| $Ca_3Ag$ | 0.090076797 | 704.6584423 |
| $Ca_2Ge$ | 0.380518224 | 701.3010812 |
| $Ca_3Cd$ | 0.147297862 | 690.8981495 |
| $Ca_3In$ | 0.209212071 | 683.823154 |
| $Ca_5Zn_3$ | 0.115732488 | 675.4375128 |

**Table S13. $\varphi_{mix}$、$\varphi_{experiment}$ and capacity of anode materials for Al batteries**

| Materials | $\varphi_{mix}$ (V) | $\varphi_{experiment}$ (V) | Capacity (mAh $g^{-1}$) |
|---|---|---|---|
| $LiAl_3$ | 0.029151 | —— | 2743.306746 |
| $MnAl_{12}$ | 0.064032 | —— | 2546.469442 |
| $Al_{41}V_4$ | 0.046711 | —— | 2515.245452 |
| $Al_{10}V$ | 0.042007 | —— | 2505.502522 |
| LiAl | 0.054478 | —— | 2369.10061 |
| $Al_{45}V_7$ | 0.058185 | —— | 2302.363024 |
| $Al_{12}Mo$ | 0.017475 | —— | 2297.709924 |
| $Al_{12}Ru$ | 0.03708 | —— | 2269.965682 |
| $Al_{23}V_4$ | 0.05439 | —— | 2242.290327 |
| $Al_4C_3$ | 0.067057 | —— | 2233.021443 |
| $MnAl_6$ | 0.14961 | —— | 2223.871236 |
| $Al_6Fe$ | 0.057223 | —— | 2214.607531 |
| $CaAl_4$ | 0.096436 | —— | 2171.993027 |
| $Li_3Al_2$ | 0.146405 | —— | 2149.223796 |
| $Mn_4Al_{19}$ | 0.186631 | —— | 2084.860939 |
| $MgAl_2$ | 0.043674 | —— | 2053.609986 |
| $Al_9Fe_2$ | 0.098739 | —— | 2040.185825 |
| $Al_9Co_2$ | 0.068759 | —— | 2005.253615 |
| $ScAl_3$ | 0.186813 | —— | 1914.989436 |
| $Li_9Al_4$ | 0.168052 | —— | 1886.58251 |
| $TiAl_3$ | 0.096483 | —— | 1871.713259 |
| $Mn_3Al_{10}$ | 0.247451 | —— | 1849.073013 |
| $Al_6Ru$ | 0.133728 | —— | 1833.732631 |
| $Al_3V$ | 0.081357 | —— | 1828.07382 |
| $Al_{13}Fe_4$ | 0.104378 | —— | 1819.695687 |
| $Al_{13}Co_4$ | 0.090816 | —— | 1781.371803 |
| $Al_{19}Co_6$ | 0.087117 | —— | 1762.727185 |
| $Al_3Fe$ | 0.039409 | —— | 1762.550168 |
| $Mg_{23}Al_{30}$ | 0.031479 | —— | 1761.822467 |
| $Al_5Mo$ | 0.054122 | —— | 1740.675763 |
| $Al_3Ni$ | 0.131113 | —— | 1726.602167 |
| $Mn_4Al_{11}$ | 0.301898 | —— | 1711.412553 |
| $CaAl_2$ | 0.166983 | —— | 1709.172497 |
| $Li_3Al$ | 0.115855 | —— | 1681.14214 |
| $Al_3Cu$ | 0.011312 | —— | 1668.611411 |
| $Ti_5Al_{11}$ | 0.12971 | —— | 1648.901307 |
| $Al_{22}Mo_5$ | 0.08278 | —— | 1647.318933 |
| $SrAl_4$ | 0.09178 | —— | 1643.93397 |
| $ScAl_2$ | 0.26472 | —— | 1624.888799 |
| $Al_{17}Mo_4$ | 0.10686 | —— | 1621.737211 |
| $Al_5Co_2$ | 0.11232 | —— | 1589.688103 |
| $TiAl_2$ | 0.144534 | —— | 1578.438786 |
| $Al_4Mo$ | 0.128021 | —— | 1576.84384 |
| MgAl | 0.019476 | —— | 1567.005284 |
| $Mn_4Al_9$ | 0.323007 | —— | 1563.594111 |
| $Ti_3Al_5$ | 0.152343 | —— | 1442.797591 |
| $Mg_{16}Al_{13}$ | 0.015939 | —— | 1412.528426 |
| $YAl_3$ | 0.192733 | —— | 1411.168731 |
| $ZrAl_3$ | 0.213988 | —— | 1400.365917 |
| $NbAl_3$ | 0.172524 | —— | 1386.817523 |
| AlP | 0.055197 | —— | 1386.689903 |
| $Al_{13}Ru_4$ | 0.224332 | —— | 1383.718343 |
| $Al_2Cu$ | 0.028328 | —— | 1367.832525 |
| $Al_8V_5$ | 0.127438 | —— | 1366.303558 |
| $Ti_2Al_3$ | 0.102214 | —— | 1364.619651 |
| $Al_3Mo$ | 0.115645 | —— | 1363.030426 |
| $Al_{12}Fe_7$ | 0.186012 | —— | 1349.384846 |
| $Al_3Ru$ | 0.155454 | —— | 1324.61432 |
| $BaAl_4$ | 0.094514 | —— | 1310.750036 |
| $Mn_5Al_8$ | 0.342505 | —— | 1310.653843 |
| $Mg_{17}Al_{12}$ | 0.011286 | —— | 1308.608639 |
| $LaAl_4$ | 0.136216 | —— | 1302.365152 |
| $Al_8Fe_5$ | 0.163874 | —— | 1298.6481 |
| $Al_3Pd$ | 0.132463 | —— | 1286.791627 |
| $Al_8Mo_3$ | 0.13201 | —— | 1276.487266 |
| $Mn_2Al_3$ | 0.369447 | —— | 1263.48639 |
| $Ca_{13}Al_{14}$ | 0.203305 | —— | 1251.875357 |
| $La_3Al_{11}$ | 0.140043 | —— | 1238.979987 |
| $Al_3Ni_2$ | 0.233141 | —— | 1215.643467 |
| $Mg_{18}Al_{11}$ | 0.038383 | —— | 1203.931134 |
| $Al_{21}Pd_8$ | 0.205004 | —— | 1190.217506 |
| $Al_3Cu_2$ | 0.086013 | —— | 1158.927696 |
| $SrAl_2$ | 0.125103 | —— | 1135.255396 |
| $Al_4Ni_3$ | 0.261928 | —— | 1131.901678 |
| $YAl_2$ | 0.274217 | —— | 1117.216932 |
| SrAl | 0.278212 | —— | 701.2704839 |
| $ZrAl_2$ | 0.309789 | —— | 1107.074965 |
| $LaAl_3$ | 0.122483 | —— | 1096.651384 |
| TiAl | 0.176711 | —— | 1073.72176 |
| $Sr_5Al_9$ | 0.137239 | —— | 1062.21565 |
| $Al_2Ru$ | 0.30109 | —— | 1036.766129 |
| AlV | 0.180603 | —— | 1031.35106 |
| $Al_2Pd$ | 0.236159 | —— | 1002.182263 |
| AlFe | 0.254356 | —— | 970.29954 |
| AlNi | 0.258996 | —— | 938.0449373 |
| AlCo | 0.184665 | —— | 935.4245475 |
| $Al_6Ge_5$ | 0.045477 | —— | 918.3190755 |
| $Zr_2Al_3$ | 0.333765 | —— | 915.3625367 |
| $AlP_2$ | 0.183706 | —— | 903.7107838 |
| AlCu | 0.129776 | —— | 887.7584836 |
| $Al_3Ru_2$ | 0.306431 | —— | 851.6881796 |
| $BaAl_2$ | 0.114369 | —— | 840.2590817 |
| $LaAl_2$ | 0.175453 | —— | 833.3799969 |
| $Al_3Pd_2$ | 0.311911 | —— | 820.6687861 |
| $Ba_{21}Al_{40}$ | 0.11498 | —— | 811.1432657 |
| $Ba_7Al_{13}$ | 0.127526 | —— | 796.2859789 |
| $Ba_3Al_5$ | 0.135727 | —— | 734.7625029 |
| SrAl | 0.145295 | —— | 701.2704839 |
| $Al_2Ni_3$ | 0.354741 | —— | 698.7128493 |
| YAl | 0.336552 | —— | 687.5556088 |
| $Sc_2Al$ | 0.499074 | —— | 687.5203176 |
| ZrAl | 0.400818 | —— | 679.8893457 |
| $Ba_7Al_{10}$ | 0.133546 | —— | 652.8016609 |
| $Al_3Ni_5$ | 0.399119 | —— | 643.9474268 |
| $Sr_8Al_7$ | 0.154883 | —— | 632.2179193 |
| AlRu | 0.44348 | —— | 627.61222 |
| $AlV_2$ | 0.327423 | —— | 623.6478202 |
| AlPd | 0.397866 | —— | 602.4422422 |
| $Ca_8Al_3$ | 0.302374 | —— | 600.3959459 |
| AlAg | 0.036846 | —— | 595.9733037 |
| $Ba_4Al_5$ | 0.15083 | —— | 587.2908927 |
| $AlFe_2$ | 0.403405 | —— | 579.5474213 |
| $Zr_5Al_4$ | 0.433721 | —— | 569.9302187 |
| $AlNi_2$ | 0.458885 | —— | 556.6815361 |
| $Zr_4Al_3$ | 0.483161 | —— | 540.7767774 |
| AlSb | 0.112488 | —— | 540.3114117 |
| $Sr_3Al_2$ | 0.119049 | —— | 507.3289902 |
| $Y_3Al_2$ | 0.39586 | —— | 496.5799766 |
| $Zr_3Al_2$ | 0.44111 | —— | 490.587115 |
| BaAl | 0.139549 | —— | 489.1211072 |
| LaAl | 0.118482 | —— | 484.4654225 |
| $Al_4Cu_9$ | 0.219619 | —— | 472.8569285 |
| $Ti_3Al$ | 0.18955 | —— | 471.1313554 |
| $Al_2Mo_3$ | 0.196393 | —— | 470.2794748 |
| $Y_5Al_3$ | 0.384864 | —— | 454.4993553 |
| $Zr_5Al_3$ | 0.45336 | —— | 448.9224788 |
| $AlV_3$ | 0.455904 | —— | 446.9600908 |
| $La_{16}Al_{13}$ | 0.150855 | —— | 406.010348 |
| $Sr_2Al$ | 0.043385 | —— | 397.4196675 |
| $AlNi_3$ | 0.485189 | —— | 395.7776235 |
| $AlCo_3$ | 0.016645 | —— | 394.3792601 |
| $Al_3Pd_5$ | 0.494156 | —— | 393.2837014 |
| $Y_2Al$ | 0.455168 | —— | 388.6331325 |
| $Al_7Te_{10}$ | 0.450013 | —— | 384.0391733 |
| $Zr_2Al$ | 0.470153 | —— | 383.7415843 |
| $Nb_2Al$ | 0.363371 | —— | 377.6751224 |
| $AlCu_3$ | 0.201491 | —— | 369.2996967 |
| $Al_2Te_3$ | 0.417764 | —— | 368.0110998 |
| $AlAg_2$ | 0.090374 | —— | 331.1126492 |
| $Al_4Ni_{15}$ | 0.492091 | —— | 325.2661326 |
| $Al_4Cu_{15}$ | 0.208495 | —— | 302.9521693 |
| $AlCu_4$ | 0.247768 | —— | 285.8347026 |
| $Nb_3Al$ | 0.357801 | —— | 262.8949951 |
| $AlAg_3$ | 0.137151 | —— | 229.2361931 |
| $Nb_4Al$ | 0.241798 | —— | 201.6201462 |
| $AlAg_4$ | 0.082615 | —— | 175.3000301 |

**Table S14. Candidate anodes for Al batteries**

| **Candidate Materials** | **Predicted $\varphi_{mix}$ (V)** | **Capacity (mA h $g^{-1}$)** |
|---|---|---|
| $LiAl_3$ | 0.029151 | 2743.306746 |
| $MnAl_{12}$ | 0.064032 | 2546.469442 |
| $Al_{41}V_4$ | 0.046711 | 2515.245452 |
| $Al_{10}V$ | 0.042007 | 2505.502522 |
| LiAl | 0.054478 | 2369.10061 |
| $Al_{45}V_7$ | 0.058185 | 2302.363024 |
| $Al_{12}Mo$ | 0.017475 | 2297.709924 |
| $Al_{12}Ru$ | 0.03708 | 2269.965682 |
| $Al_{23}V_4$ | 0.05439 | 2242.290327 |
| $Al_4C_3$ | 0.067057 | 2233.021443 |
| $MnAl_6$ | 0.14961 | 2223.871236 |
| $Al_6Fe$ | 0.057223 | 2214.607531 |
| $CaAl_4$ | 0.096436 | 2171.993027 |
| $Li_3Al_2$ | 0.146405 | 2149.223796 |
| $Mn_4Al_{19}$ | 0.186631 | 2084.860939 |
| $MgAl_2$ | 0.043674 | 2053.609986 |
| $Al_9Fe_2$ | 0.098739 | 2040.185825 |
| $Al_9Co_2$ | 0.068759 | 2005.253615 |
| $ScAl_3$ | 0.186813 | 1914.989436 |
| $Li_9Al_4$ | 0.168052 | 1886.58251 |
| $TiAl_3$ | 0.096483 | 1871.713259 |